\documentclass[11pt]{article}
\usepackage{a4wide,graphicx,epsfig}
\def\question#1 {~\\ {\bf\it #1 }\\}
\def\question#1 {}
\newcommand{\A}{{\mathcal A}}
\newcommand{\p}{{\mathcal A}}
\newcommand{\be}{\begin{equation}}
\newcommand{\ee}{\end{equation}}
\newcommand{\ba}{\begin{eqnarray}}
\newcommand{\ea}{\end{eqnarray}}
\newcommand{\bb}{\begin{thebibliography}}
\newcommand{\eb}{\end{thebibliography}}
\newcommand{\ci}[1]{\cite{#1}}
\newcommand{\bi}[1]{\bibitem{#1}}
\newcommand{\lab}[1]{\label{#1}}
\newcommand{\ua}{\sigma^{(\uparrow)}}

\newcommand{\da}{\sigma^{(\downarrow)}    }

\newcommand{\ahnf}{${\mathcal A}^h_{nf}$}
\newcommand{\acsf}{${\mathcal A}^{em}_{sf}$}

\begin{document}

\begin{center}
{ \large\bf High-energy hadron spin-flip amplitude at small momentum transfer 
 and new $A_N$ data from RHIC }  \\
\vskip 1cm

J.-R. Cudell\footnote{Institut de Physique, 
 B\^at. B5a, Universit\'e de Li\`ege, Sart Tilman, B4000
  Li\`ege, Belgium, e-mail: J.R.Cudell@ulg.ac.be }, 
E. Predazzi\footnote{Dipartimento di Fisica Teorica - Universit\`{a} di Torino
   and Sezione INFN di Torino, Italy, e-mail: predazzi@to.infn.it.}, 
 and O.V. Selyugin$^{1,}$\footnote{on leave from the Bogoliubov
 Laboratory of Theoretical Physics, JINR, 141980, Dubna, Moscow Region,
 Russia, e-mail:selugin@qcd.theo.phys.ulg.ac.be.}\\
\end{center}
\vskip 1cm
\begin{quote}
\centerline{\bf Abstract}
{\small\noindent   In the case of elastic high-energy hadron-hadron 
scattering,  
 the impact of the large-distance contributions on the behaviour 
  of the slopes of the spin-non-flip and of the spin-flip amplitudes 
  is analysed.
  It is shown that the long tail
   of the hadronic potential
   in  impact parameter space
   leads to a value of the slope 
   of the reduced spin-flip amplitude larger than that of
   the spin-non-flip amplitude.
  This effect
  is taken into account in the calculation of the analysing power
  in  proton-nucleus reactions at high energies. It is shown that
  the preliminary measurement of $A_N$ for $p ^{12}C$  obtained by the E950
Collaboration indeed favour a spin-flip-amplitude with a large slope.
 Predictions for $A_N$ at $p_L = 250 \ $~GeV/$c$ are given.\\}
\end{quote}

\vskip 2cm

\section{Introduction}

Diffractive polarised experiments open a new window on the
spin properties of QCD at large distances. In particular,
the recent data from RHIC and HERA indicate 
that, even at high energy, the hadronic amplitude has 
a significant spin-flip contribution, ${\mathcal{A}}^{h}_{sf}$,
which remains proportional to the spin-non-flip part, \ahnf, 
as energy is increased.
There were many observations of spin effects at high energies and
at fixed momentum transfers.
 Several
attempts to extract the spin-flip amplitude from the experimental
data  show that the ratio of spin-flip to spin-non-flip
amplitudes can be non-negligible and may be only slightly dependent on energy
\cite{soffer,akch,sel-pl}.
   Thus, the diffractive polarised experiments at HERA and RHIC
 allow one to study spin properties of quark-pomeron and
 proton-pomeron vertices
 and  to search  for a possible  odderon contribution.
      This provides an important test of the
 spin properties of QCD at large distances.
In all of these cases,  pomeron exchange
is expected to contribute to the observed spin effects at some level
 \cite{martini}.

In the framework of perturbative QCD,
 it was shown that
the analysing power of hadron-hadron
scattering  can be non negligible and proportional to
 the hadron mass \cite{ter}.  Hence, one would
expect a large analysing power for moderate $p_{t}^{2}$,
where the spin-flip amplitudes are presumably relevant
 for  diffractive processes.
 When large-distance
contributions are considered, one obtains  a more complicated
spin structure for the pomeron coupling.
For example, the spin-flip amplitude has been estimated in the QCD Born
approximation by  the non-relativistic quark model for the
nucleon wave function \cite{kopel} in the case where the nucleon
contains a dynamically enhanced component with a compact di-quark.
The spin-flip part of the scattering
amplitude can be determined by the hadron wave function for the
pomeron-hadron couplings or by the gluon-loop corrections for the
quark-pomeron coupling \cite{gol-pl}.  As a result, the spin asymmetries
that appear have a weak energy dependence as $s \to \infty$.
Additional spin-flip contributions to the quark-pomeron vertex may
also have their origins in instantons (see {\it e.g.} \cite{fo,do}).

 The inclusion in the analysis of the experimental data on spin-correlation
parameters does not simplify the task.
 In the general case,
the  form of the analysing power, $A_N$, and
 the position of its maximum, 
 depend on the parameters of the elastic scattering
 amplitude, {\it i.e.}  $\sigma_{tot}$,  $\rho(s,t)$, 
the Coulomb-nucleon interference
 phase  $\varphi_{cn}(s,t)$
 and the elastic slope $B(s,t)$.
 For the definition of new effects at small angles,
  and especially in the region of the diffraction minimum,
  one must  know the effects of the Coulomb-hadron interference
 with sufficiently high accuracy.
  The Coulomb-hadron phase was calculated
 in the entire diffraction domain taking into account  the form factors
 of the  nucleons \cite{prd-sum}.
  Some polarisation effects connected with the Coulomb-hadron
   interference, including some possible odderon contribution,
   were also calculated \cite{z00}.

   The dependence of the hadron spin-flip amplitude on momentum transfer
  at small angles is tightly connected with the basic structure of the
  hadrons at large distances. We shall show that
  the slope of the  ``reduced'' hadron spin-flip amplitude
  (the hadron spin-flip amplitude without the kinematic factor $\sqrt{|t|}$)
  can be larger
  than the slope of the hadron spin-non-flip amplitude,
  as was observed long ago \cite{predaz,wak}.
  This leads to small effects in the differential hadron cross section
  and  in  the real part of the hadron non-flip amplitude \cite{sel-sl}.

The new RHIC fixed-target data, from E950, consist 
in measurements of the analysing power 
\be A_N(t)={\ua-\da\over \ua+\da}\ee
for momentum transfer $0 \leq |t| \leq 0.05$~GeV$^2$,
for a polarised $p$ beam hitting a (spin-0) $^{12}C$. In this 
region of $t$, 
the electromagnetic amplitude is of the same order of magnitude 
as the hadronic amplitude, and the interference of the imaginary part 
of \ahnf $ \ $ with the spin-flip part of the 
electromagnetic amplitude \acsf $ \ $ leads to a peak in the analysing 
power $A_N$,
usually referred to as the Coulomb-Nuclear Interference (CNI) effect
\cite{schwinger,bib14,lead}. This effect was observed in the data from
\ci{akex}, but the errors were too big to draw any conclusion 
on the hadron spin-flip amplitude.

The first RHIC measurements at $p_L = 22 $~GeV/$c$ \cite{an22}  
in $p ^{12}C$ scattering indicated however that
$A_N$ may change sign already at very small momentum transfer.
Such a behaviour cannot be described by the CNI effect alone. Indeed,
fits to the data \cite{kt} give  
for 
\ba r_5&=& \lim\limits_{t \to 0}{ \ \tilde{ \mathcal A}^h}_{sf}
                                      /{Im(\mathcal A}^h_{nf})
\equiv R+iI:\label{r5}\\
     R&=&0.088 \pm 0.058;
     \ \ \ I=-0.161 \pm 0.226  
\ea

\section{Ratio of slopes of spin-flip and spin non-flip amplitudes}

As usual, 
$$ \tilde{\A}_{sf}(s,t) \equiv  2\ m_p\ \A_{sf}(s,t)/ \sqrt{|t|} $$
 is the  ``reduced'' spin-flip amplitude,
factoring out trivial kinematic factors.
The large error on $Im(r_5)$ unfortunately 
leads to a high uncertainty on the size of 
the hadronic spin-flip amplitude. 
%???
% Note that the large errors in the size of the coefficients come not only 
% from the large experimental errors, but  also 
%  from the form of the basic
% formula (34) \cite{kt} in which the last term with  the difference
% $Re r_5^{pA}- \rho_{pA} Im r_{5}^{pA}$
% reflected the hadron-hadron
%  interference equals a zero by the definitions used authors. 

For spin 1/2 scattering, the total helicity amplitudes can be decomposed
in sub-amplitudes describing the ways the two spins can be changed during the
collision:
$$ \Phi_i(s,t) = \phi^h_{i}(s,t)
        + \phi_{i}^{em}(t) \exp[i \alpha_{em} \varphi_{cn}(s,t)], i=1,\ 5$$ 
 where $\phi^h_{i}(s,t)$ represents the pure strong interaction of hadrons,
  $\phi_{i}^{em}(t)$  represents the electromagnetic interaction of hadrons,
  $\alpha_{em}=1/137$ is the electromagnetic constant,
   and
  $\varphi_{cn}(s,t)$ is the electromagnetic-hadron interference phase factor.
  So, to determine the hadron spin-flip amplitude
 at small angles, one should take
  into account all electromagnetic and all 
  electromagnetic-hadronic  interference effects.

 In this paper, as we shall be interested in scalar targets,
we define the spin-non-flip amplitudes as
  ${\A}^{h}_{nf}(s,t)
    = (\phi^h_{1}(s,t) + \phi^h_{3}(s,t))/(2s)$ for the hadronic one,
  and
 $\A^{c}_{nf}(s,t)
    = (\phi^{em}_{1}(s,t) + \phi^{em}_{3}(s,t))/(2s)$ 
 for the electromagnetic one. 
 Taking into account the Coulomb-nuclear phase $\varphi_{cn}$, we
 obtain $Im \A_{nf}^{c} \approx \alpha_{em} \varphi_{cn} |\A_{nf}^{c}|$.
The   ``reduced''    spin-flip amplitudes are denoted as
  $\tilde{\A^{h}_{sf}}(s,t) =  \phi^{h}_{5}(s,t)/(s \sqrt{|t|})$  and
  $\tilde{\A^{c}_{sf}}(s,t) =  \phi^{em}_{5}(s,t)/(s \sqrt{|t|})$.

  As usual, we define the slopes $B_i$ of the scattering amplitudes as
   the derivatives of the logarithm of the amplitudes with respect to $t$.
     For an exponential form of the amplitudes,  this  coincides
    with the standard slope of the differential cross sections divided by $2$.
  If we define the forms of the separate  hadron scattering
  amplitude as:
\ba
Im \ \A_{nf}(s,t) &\sim& \exp(B_{1}^{+} \ t), \ \ \
  Re \ \A_{nf}(s,t) \sim \exp(B_{2}^{+} \ t), \nonumber \\
 Im  \tilde{\A_{sf}}(s,t) & \sim &
  \ \exp(B_{1}^{-} \ t), \ \ \
  Re \tilde{\A_{sf}}(s,t) \  
\sim  \ \exp(B_{2}^{-} \ t),
\ea
 then, at small $t$ (in $[0, 0.1]$ GeV$^2$), almost all
  phenomenological analyses assume
$ B_{1}^{+} \ \approx \ B_{2}^{+} \ \approx \
    B_{1}^{-} \ \approx \ B_{2}^{-} . $
    To obtain this, we can take the eikonal representation for the scattering
   amplitude
\ba
\phi^{h}_{1}(s,t) & =& - i p \int_{0}^{\infty} \ \rho \ d\rho
 \ J_{0}(\rho q)
 [e^{\chi_{0}(s,\rho)} \  - \ 1 ], \nonumber \\
\phi^{h}_{5}(s,t) &=& - i p \int_{0}^{\infty} \ \rho \ d\rho
 \ J_{1}(\rho q) \ \chi_{0}(s,\rho) \ e^{\chi_{0}(s,\rho)},
\ea
where $p$ is the momentum in the centre-of-mass frame, $q=\sqrt{-t}$ is
the momentum transfer, and 
$\chi_{i}(s,\rho)$ are the eikonal phases in impact-parameter ($\rho$) 
space coming from the spin-non-flip ($i=1$) and
spin-flip interaction ($i=2$) potentials $V_i(\rho,z)$.
If the potentials $V_{i}$ are
   assumed to have the same Gaussian form,
  in the first Born approximation, 
  $\phi^{h}_{1}$ and $\hat{\phi_h}^{5}$
   will also have the same Gaussian form
\ba
 \phi^{h}_{1}(s,t) & \sim &  \int_{0}^{\infty} \ \rho \ d\rho
 \ J_{0}(\rho q) \ e^{- \rho^{2}/2R^{2} } \ = \ R^2 \  e^{R^{2} t/2}, 
                                \nonumber \\ 
 \hat\phi^{h}_{5}(s,t) & \sim & {1\over q}\int_{0}^{\infty} \ \rho^2 \ d\rho
 \ J_{1}(\rho q) \  
   \ e^{-\rho^2/(2R^2) } \ =  \ R^4 \ e^{ R^{2} t/2}  . \lab{f5a}
\ea
  In this special case, the slopes of
 the  spin-flip and  ``residual''spin-non-flip amplitudes are
  indeed the same.

  However, a Gaussian form of the potential
 is at best adequate to represent the central part of the
   hadronic  interaction. This form cuts off the Bessel function
  and the contributions at large distances.
   If we expand the $J_i (x)$ at small $x$ and truncate the series
   at order $x^2$, we obtain
\ba
J_{0}(x) \simeq  1  -  (x/2)^2 ; \ \ \ {\rm and} \ \ \
  2 \ J_{1}/x \ \simeq \ (1 \ - 0.5 \ (x/2)^2) , \lab{sbess}
\ea
and the corresponding integrals 
\ba
 \! \! \! \!  \phi^{h}_{1}(s,t) & \sim &  \int_{0}^{\infty} \ \rho \ d\rho
 \ \left(1- \rho^2 \frac{q^2}{4}\right) \ e^{-\rho^2/2R^2} \ 
                                  \approx \ R^2 e^{-R^{2}~ q^{2}/2}, 
                                \nonumber \\ 
  \! \! \! \!   \hat\phi^{h}_{5}(s,t) & \sim & 
  \frac{1}{q} \ \int_{0}^{\infty} \ \rho^2 \ d\rho
 \ \rho \ \frac{q}{2} \ \left(1- \rho^2 \frac{q^2}{8} \right) \  
   \ e^{- \rho^2/2R^2 } \ 
     \approx  \ \ R^4 \ e^{- R^{2} q^{2}/2}  . \lab{f6b}
\ea
   still have the same behaviour  over $q^2$ \cite{tur1}.  
   So, the integral representation  for spin-flip and spin-non-flip amplitudes
 will be the same as in (\ref{f5a}).

  If, however, the potential (or the corresponding eikonal)
 has a long (exponential or power) tail
  in impact parameter,  
  the approximation (\ref{sbess}) for the Bessel functions
  does not lead to correct results and one has to perform
 the full integration.

  Let us examine the contribution of large distances.
  The Hankel asymptotics of the Bessel functions at large distances
  \cite{sprav} are
\ba
 J_{\nu}(z) \ & = & \  \sqrt{2/ \pi z} \ [ P(\nu , z) \ \cos{\chi(\nu,z)} \ -
 \  Q(\nu,z) \  \sin{\chi(\nu, z)} ], \nonumber \\
    P(\nu,z) \ &  \sim & \ \sum_{k=0}^{\infty} \  (-1)^{k} \  
   \frac{(\nu, \ 2k)}{(2z)^{2k} } ,   \nonumber \\ 
%      \frac{[(4k-1)!!]^2}{(2k)! \ (8z)^{2k}},   \nonumber \\ 
  Q(\nu,z) \  & \sim & \ \sum_{k=0}^{\infty} \ (-1)^{k} \  
          \frac{(\nu, \ 2k+1)}{(2z)^{2k+1} } ,   \nonumber \\ 
%      \frac{[(4k-1)!!] \ [4k+1)!!]}{(2k!) \ (8z)^{2k+1}}; \nonumber \\  
   \chi(\nu,z) & = & z- (\nu/2 +1/4).  \lab{asbes} 
\ea
  This gives for $\nu =0,~1$

\ba
 J_{0}(x) \ = \   \sqrt{2/ (\pi x)} [P_{0}(x) \cos{(x-\pi/4)}-Q_{0}(x), 
  \sin{(x-\pi/4)}] \ \ \ \nonumber \\
  J_{1}(x) \ = \ \sqrt{2/ (\pi x)} [P_{1}(x) \cos{(x- 3\pi/4)} - Q_{1}(x) 
  \sin{(x- 3\pi/4)} ] , \lab{sbes}
\ea
 with
\ba
 P_{0}(x) &\approx&  1-0.0703125/x^2+0.1121521/x^4 + ...; \nonumber \\
 Q_{0}(x) &\approx& -0.125/x +0.073242188/x^3 + ... ; \nonumber \\
  P_{1}(x)& \approx&  1 + 0.1171875/x^2 -0.144195557/x^4 + ... ;  \nonumber \\
  Q_{1}(x) &\approx& 0.375/x-0.10253906/x^3 + ... . 
\ea
 From this, we obtain:
\ba
\! \! \! \! \! \! \sqrt{\pi x } J_{0}(x)  \  \approx  \   
  (1 - \frac{0.125}{x} &-& 
      \frac{0.07}{x^2}) \cos{x}
  +(1+ \frac{0.125}{x} - \frac{0.07}{x^2}) \sin{x};  \nonumber \\
 \! \! \! \! \! \!   \sqrt{\pi x} J_{1}(x)  \ \approx  
  (1 + \frac{0.375}{x} &+& \frac{0.117}{x^2}) \sin{x}
    -(1- \frac{0.375}{x} + \frac{0.117}{x^2}) \cos{x}. \lab{rowj2}
\ea
   The leading behaviour at large  $x$  
 will thus  be proportional  to $1/\sqrt{q \rho}$.

 Let us calculate the corresponding  integrals in the case 
  of large distances
\ba
\phi^{h}_{1}(s,t) & \sim&  \int_{1}^{\infty} \ 
\rho^2/\sqrt{q \ \rho} \  \ \exp[-\rho^2/(2 R^2)] \ d\rho  \nonumber \\
 & = & \ 1/\sqrt{2 \ q} \  \ R^{3/2} \ \gamma[3/4;1/(2R^2)]  \nonumber \\
\phi^{h}_{5}(s,t)/q  & \sim & 1/q \ \int_{0}^{\infty} \ 
  \rho^2/\sqrt{q \ \rho} \ \ \exp[-\rho^2/(2 R^2)] \ d\rho \nonumber \\
  & = &  1/(q \ \sqrt{q}) \ \  2^{1/4} \  \ R^{5/2} \ \ \gamma[5/4;1/(2R^2)]   
    . \lab{fasj2}
\ea
  where $\gamma[a,z]$ is the incomplete Gamma function.
 We see that the exponential asymptotics of  both representations 
are the same, but the
 reduced  spin-flip  amplitude has an additional $q^{3/2}$ in the denominator
 and hence it has a larger effective slope then the spin-non-flip amplitude. 
%\ba 
% \Gamma(a,z) = \Gamma(a)- \gamma(a x)
%\ea
  Slightly more complicated calculations, keeping the
  $O(1/x)$ in (\ref{rowj2}), lead to 
   practically  the same results:
\ba
 \phi^{h}_{1}(s,t) &\sim&  1/q^2  \int_{0}^{\infty}  \sqrt{x} 
\left[\left(1- {0.125\over x}\right) \cos{x} 
  +\left(1+{0.125\over x}\right) \sin{x}\right]  
 e^{-x^2/(2 R^2 q^2)}  dx \nonumber\\ 
 &\approx& {R\over q} \ _{1}F_{1}(3/4,1/2,-q^2 R^2/2),  \nonumber \\
 \frac{\phi^{h}_{5}(s,t)}{q}  
   &\sim&  \frac{1}{q^4}  \int_{0}^{\infty}  \ x^{3/2} \ 
   \left[\left({0.375\over x}-1\right) \cos{x}  
+\left(1- \frac{0.375}{x}\right) \sin{x}\right]  
                   e^{-x^2/(2 R^2 q^2)} \ dx   \nonumber \\    
& \approx& {R^{3/2}\over q^{5/2}} \ _{1}F_{1}(3/4,1/2,-q^2 R^2/2), 
\ea
 Again, the additional $q^{3/2}$ in the denominator leads to a larger
  slope for the residual spin-flip-amplitude. 
  So, despite the fact that the integrals have the same 
   exponential  behaviour asymptotically,
   the  additional inverse power of $q$ leads to  
   a larger  effective slope for the 
  residual spin-flip amplitude, although we take a Gaussian representation
  in impact parameter. 
 
  These  results  can be confirmed by a numerical calculation of
   the relative contributions of the large distances. 
  We calculate the scattering amplitude in the Born 
  approximation in the cases
  of exponential and Gaussian form factors in impact parameter representations
  as a function of the upper limit b of the corresponding integral
\ba
\phi^{h}_{1}(t) &\sim&  \int_{0}^{\rm b} \ \rho \ d\rho
 \ J_{0}(\rho \Delta) f_{n} , \ \ 
%   \nonumber \\
\phi^{h}_{5}(t)/q \sim  \int_{0}^{\rm b} \ \rho^2 \ d\rho
 \ J_{1}(\rho \Delta) \  f_{n}  . \lab{f5ab}
\ea
 with $f_{n}= \exp{[-(\rho/5)^n]}$, and $n=1,\ 2$. 
  We then  calculate the ratio of the slopes of these two amplitudes 
   $R_{BB} =B^{sf}/B^{nf}$ as a function of b for these two values of $n$.
  The result is shown in Fig.~1.
  We  see that at small impact parameter the value of $R_{BB}$ is 
  practically the same in both cases and depends weakly on   the value of  
  b. However, at large distances, the behaviour of $R_{BB}$ is different.
  In the case of the Gaussian form factor,  the value of $R_{BB}$
  reaches its asymptotic value ($=1$) quickly.
   But in the case of the exponential behaviour,
  the value $R_{BB}$ reaches  its limit $R_{BB}=1.7$ only at large distances.
  These calculations confirm our analytical analysis of the asymptotic
  behaviour of these integrals at large distances.

\begin{figure}
\epsfysize=6.cm
\epsfxsize=8.5cm
\vglue -1cm
\centerline{\epsfbox{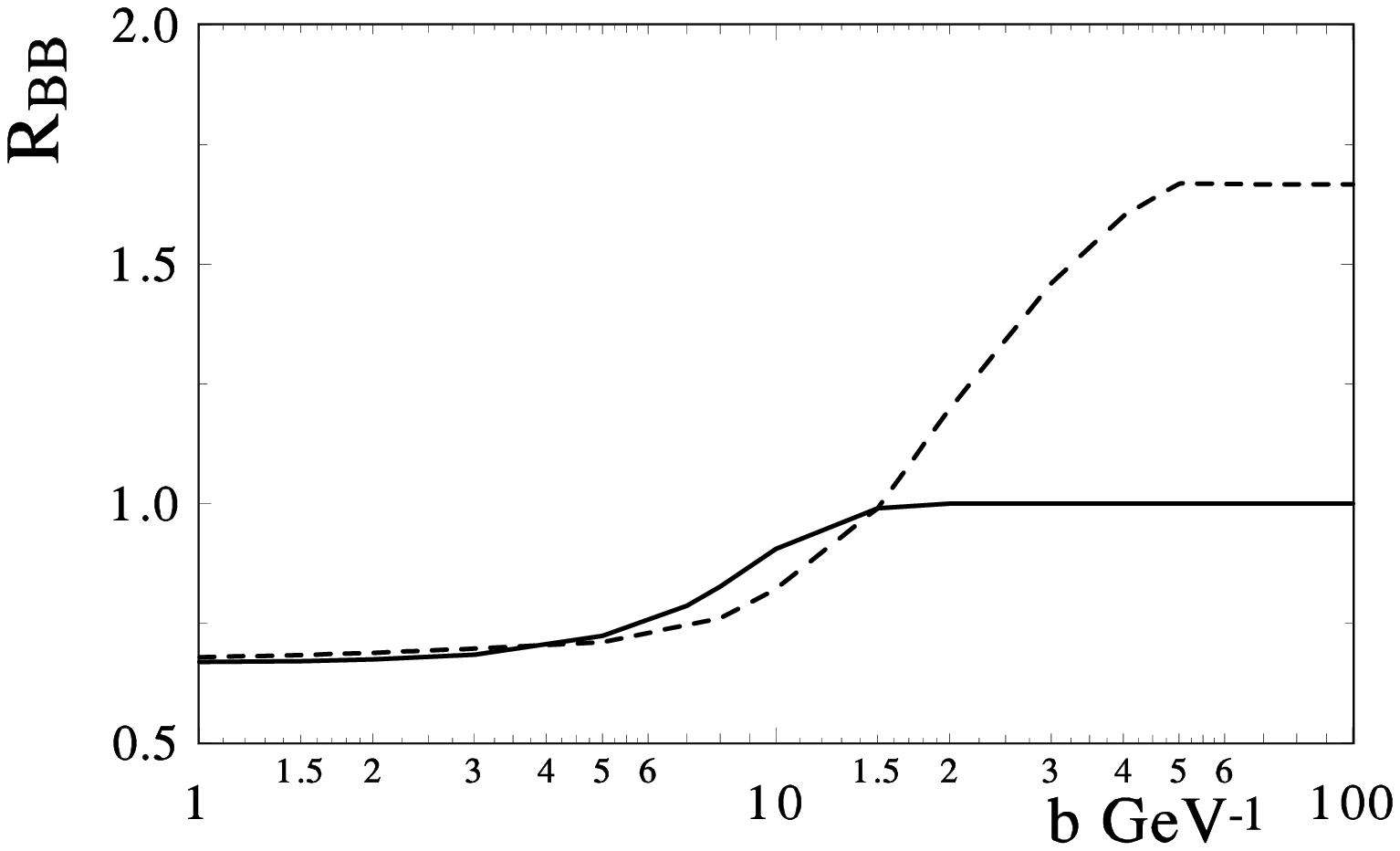}}
%\vglue -1cm
%\centerline{\epsfbox{c12bhxb.ps}}
\caption{ The ratio of the effective slopes - $R_{BB}$ for the cases $n=1$
 (dashed line) and $n=2$ (solid line) as a function of
 the upper bound of the integrals
  b.}
\end{figure}

  The first observation that the slopes don't coincide
 was made in \cite{predaz}.
  It was found from the analysis of the
 $\pi^{\pm} p \rightarrow \ \pi^{\pm}p $ and
  $pp \rightarrow \ pp $
reactions
  at $p_L \ = \ 20 - 30 \ {\rm GeV}/c $
  that the slope of the ``residual'' spin-flip amplitude is about
   twice as large as
  the slope of the spin-non-flip amplitude. This conclusion can also
  be reached
  from the phenomenological analysis carried out in \cite{wak} of
  spin correlation parameters in elastic proton-proton scattering
  at $p_L \ = \ 6 \ $GeV$/c$.

  In \cite{tur1}, it was shown that in the case of an exponential tail
  for the  potentials
$ \chi_{i}(b,s) \ \sim \ H \ e^{- a \ \rho}, $
 and using the standard integral representation
\ba
\int_{0}^{\infty} \ x^{\alpha-1} \  \exp(-p \ x) J_{\nu}(cx)\ dx \
   = \ I_{\nu}^{\alpha}, \nonumber
\ea
with $
 I_{\nu}^{\nu+2} \ = \ 2 p \ (2c)^{\nu} \ 
\Gamma (\nu + 3/2)/[\sqrt{\pi} (p^{2}+c^{2})^{3/2}], 
$
 one  obtains
%{\Large
\ba
 \A_{nf} (s,t) & \sim &  \frac{1}{a\sqrt{a^2+q^2}} \ e^{-B q^2}, \ \  
%\nonumber \\
\tilde{\A_{sf}}(s,t) \sim
 \ \frac{3 \ a  \ B^{2}}{ \sqrt{a^2+q^2}} \ \ e^{-2 \ B q^2}.
\ea     
  In this case, one can see that
  the slope of the ``residual'' spin-flip amplitude exceeds the slope
 of the spin-non-flip amplitudes by a factor of two.

%%%%%%%%%%%%%%%%%%%%%%%%%%%%%%%%%
   It is interesting to note that the derivative relations for the helicity
  amplitudes with
  $t \not= 0$ and 
 $\{ \lambda_i \} = \lambda_c, \lambda_d, \lambda_a, \lambda_b \ \ $,   
  $\Delta \lambda = | \lambda_c - \lambda_d - \lambda_a + \lambda_b | \  $
  for spin-dependent amplitudes, 
  carefully examined in \cite{shremp1}, 
\ba
 F_{\lambda_i}(s,t) = C_{\lambda_i}(s) \left(\sqrt{-t}\right)^{\Delta \lambda }
  \left(\frac{1}{\sqrt{-t}} \frac{\partial}{\partial \sqrt{-t}}\right)^{\Delta \lambda} 
  F_{\Delta \lambda=0}(s,t)   \lab{ddr}
\ea
lead to the same results. In the case of a Gaussian form for the 
spin-non-flip amplitude, using (\ref{ddr}), we obtain the same slopes for
  the spin-flip and spin-non-flip amplitudes. But if we choose another $t$ 
dependence for the
  spin-non-flip amplitude, for example that given by 
  the hadronic form factor 
 $ \Lambda^2/ (\Lambda^2 - t)^{n} $,
 the spin-flip amplitude  is then given by
\ba
   F_{1/2,-1/2}(s,t) = C_{1/2,-1/2}(s) \sqrt{-t}
   \frac{\Lambda^{2}}{ (\Lambda^2 - t)^{n+1} }. 
\ea
  In the case $n=1$, the  slope becomes about twice as large.
  The same result was obtained in the impact-parameter representation
  in \cite{tur1}. Of course, the derivative relations
   at non-asymptotic energy should be used
  with care, as was done for the derivative 
  dispersion relations  for forward scattering \cite{martyn}.

%%%%%%%%%%%%%%%%%%%%%%%%%%%%%%%%%%%%%%%%%%%%%%%

  Hence, a  long-tail hadronic potential
 implies a
 significant difference in the
  slopes of the ``residual'' spin-flip and of the spin-non-flip amplitudes.
   Note also that the procedure
  of  eikonalisation will lead to a further increase
  of the difference between these two slopes.

\section{The analysing power in proton-nucleus scattering}

  The above results can be used in the description of the analysing power at
  small momentum transfer. In the case of hadron-hadron scattering
  at large energy, the experimental data are scarce. The most famous
  experiment on proton-proton scattering at $p_L = 200 \ $GeV$/c$ 
  has large errors, and the analysis of \cite{akch} concludes
  that the hadron spin-flip amplitude contributes very little. 
  Of course, it will be very interesting obtain further measurements
from the RHIC PP2PP collaboration. 
At present, we only have preliminary experimental data
  on $A_N$ in proton-Carbon elastic scattering. Despite the fact
 that these data have bad normalisation conditions, 
 the form of the analysing power is already
   very interesting.  

  For $p ^{12}C$ scattering, the elastic and total cross sections, and 
 the analysing power 
$A_N$, are given by
\ba
d\sigma/dt  &=&  \phantom{-} \pi \left(|\A_{nf}|^2+ |\A_{sf}|^2\right), 
 \nonumber\\
\sigma_{tot}&=& \phantom{-}4 \pi\ Im(\A_{nf}),\label{an}\\
A_N \ d\sigma/dt  &=&   - 2\pi\ 
                   Im\left( \A_{nf} \A_{sf}^{*}\right). \nonumber
\ea
 
Each term includes a hadronic and an electromagnetic contribution:
$
 \A_i(s,t) = \p^h_{i}(s,t) 
        + \p_{i}^{em}(t) e^{i\delta}, (i=nf, sf),
$
where $\p^h_{i}(s,t)$ describes the strong interaction of $p$ with $ ^{12}C$,
and $\p_{i}^{em}(t)$  their  electromagnetic interaction.
$\alpha_{em}$ is the electromagnetic fine structure constant, and
the Coulomb-hadron phase $\delta$ is given by
$\delta=Z\alpha_{em} \varphi_{CN}$ 
with $Z$ the charge of the nucleus, and $\varphi_{CN}$
the Coulomb-nuclear phase \cite{prd-sum}.
The electromagnetic part of the scattering amplitude can be written 
as
\ba
\A^{em}_{nf}&=& {2 \alpha_{em} \ Z\over t} \ F^{^{12}C}_{em} F^{p}_{em1},
 \\ \nonumber
\A^{em}_{sf}&=& - {\alpha_{em} \ Z\over m_p \sqrt{|t|}} \ 
                           F^{^{12}C}_{em} F^{p}_{em2},
\ea
where 
$ F^{p}_{em1}$ and $ F^{p}_{em2}$ are
the electromagnetic form factors of the proton, and
$F^{^{12}C}_{em}$ that of $^{12}C$.
We use 
\ba 
F^{p}_{em1}&=& {4 m_p^2-t (\kappa_p+1)\over (4 m_p^2-t)(1-t/0.71)^2} , \\ 
F^{p}_{em2}&=& {4 m_p^2\kappa_p\over (4 m_p^2-t)(1-t/0.71)^2} , 
\ea
 where $m_p$ is the mass of the proton and $\kappa_p$ its anomalous 
magnetic moment.
We obtain $F^{^{12}C}_{em}$
from the electromagnetic density of the nucleus 
\ba
D(r) = D_0 \ \left[1+ \tilde{\alpha}\left({r\over a}\right)^2
\right]e^{-\left({r\over a}\right)^2}.
\ea
$\tilde{\alpha}=1.07$ and $a=1.7$ fm give the best description
 of the data \cite{jansen} in the small-$|t|$ region,
and produce a zero of $F^{^{12}C}_{em}$ at $|t| = 0.130 $~GeV$^2$. 
We also calculated $F^{^{12}C}_{em}$ by integration of the nuclear 
form factor given by a sum of Gaussians \cite{data35} and obtained 
practically the same result with the zero now at  $|t| = 0.133$ 
GeV$^2$. 

We now need to model the strong-interaction parts of the amplitude.
Isoscalar targets such as $^{12}C$ simplify the calculation as they 
suppress the contribution of the isovector reggeons $\rho$ and $a_2$, 
by some power of the atomic number. Also, as $^{12}C$ is spin 0,
there are only two independent helicity amplitudes: proton spin flip 
and proton spin non flip.
However, nuclear targets lead to large theoretical uncertainties
because of the difficulties linked to nuclear structure, and 
 of the lack of high-energy proton-nucleus scattering 
experiments (see, for example, \cite{harr}).

For the $t$ dependence, as we shall be concerned with the small-$t$
region, we shall
assumed that they are well approximated by falling exponentials.
 The slope parameter $B(s,t)/2$ is then the derivative 
 of the logarithm of the amplitude with respect to $t$.  
 If one considers only one contribution to the amplitude, 
 this corresponds with 
 the slope $B(s,t)$ of the differential cross section.

  Specific questions appear when we consider the energy dependence of the
  spin-non-flip and spin-flip amplitudes and their phase, or their 
  ratios $\rho$ of real to imaginary parts.   

  For the latter, the situation is not settled.
  Firstly, we may expect the size of  $\rho^{pA}$ 
  to be $\rho^{pA} = \rho^{pp}/2$,  as the 
  $a_2$ and $\rho$ contributions 
  decrease in the nucleus.   
   Secondly, experimental data 
on  proton-deuteron \cite{pd} however show 
   practically the same size for $\rho^{pD}$ and $\rho^{pp}$, 
   from which we might conclude\footnote{ The analysis based 
on partial wave amplitudes 
  and dispersion relations
  \cite{kroll} leads qualitatively to the same result. Note that this
  analysis works well at low energies, but that 
  it has some problems at high energies. 
For example, the values of $\rho (p\bar{p})$ from
  \cite{kroll2} seem to conflict with recent data \cite{pbp}. This may be
  due to the fact that the data on which one must rely for such a study
  sometimes contradict each other, mainly because
  $\rho$ is not a direct observable, but requires some  theoretical input which
varies from one experiment to the other.} that  $|\rho_{pn}| \leq |\rho_{pp}|$.
   Thirdly, ref.~\cite{bujak} shows that 
   $|\rho_{pp}| > |\rho_{pD}| > |\rho_{pHe}|$, and
   leads to the conclusion that the size of $\rho$ depends 
  on the atomic number   with 
  $\rho_{pA}  \simeq \rho_{pp}/A $. 
   In this case, we obtain for  $p ^{12}C$-scattering
  at large energies  $\rho \approx 0$.  
Finally,
the data from \cite{shiz,selex} indicate that it is very likely  
 that $\rho^{pA} \geq 0$.

  In this paper, we choose an intermediate variant: 
   \ba\rho_{p ^{12}C} \approx \rho_{pp}/2,\ea which
  corresponds to $\rho_{pA} = \rho_{pp} A^{-1/3}$.
 We emphasise that we do not know the energy dependence of $\rho^{pA}$, 
but because the $\rho$ and $a_2$ trajectories are suppressed 
 in proton-Carbon scattering, and
because they contribute negatively, it
must be larger  than in the $pp$ case, where it is about $-0.1$ in 
this energy region.
    
   The energy dependence of the 
   asymptotic  spin-flip amplitude is also far from decided. 
   As mentioned above, it is possible that, in high-energy hadron-hadron 
   scattering, the ratio of the spin-non-flip to the 
   spin-flip amplitude decreases slowly with energy. 
   In this case, if we take only the asymptotic 
   part of the spin-flip amplitude into account, 
   we cannot make its real part proportional
   to $\rho_{pp}$. In the following, we shall see that 
   imposing proportionality to $\rho_{pp}$ leads to
   a strong energy dependence for the spin-flip amplitude.
    
The above considerations lead to the following form for
the hadron spin-non-flip amplitude:
  \ba
\A^{pA}_{nf}(s,t) &=& (1+\rho^{pA}) 
{\sigma_{tot}^{pA}(s)\over 4\pi} \exp\left({B^{+}\over 2}t\right) .
\label{f-h}
\ea
  The slope and $\rho$ parameter are assumed to be proportional to 
  their  values in $pp$-scattering, which  we take as in  \cite{selex} :    
\ba
B^+_{pp}(s)& =& 11.13 -  6.21/\sqrt{p_{L}} - 0.3 \ \ln{p_{L}}, \nonumber \\
\rho^{pp}(s)& =& 6.8/p_{L}^{0.742} -  6.6/p_{L}^{0.599}+0.124 .
\ea

The total cross sections were chosen as \ba\sigma_{tot}^{pA}= R_{C/p}(s)
\sigma_{tot}^{pp}\ea. The $pp$ total cross section is obtained from
the best form 
 of \cite{prd-comp}, which works well in this energy region:
\ba
\sigma_{tot}^{pp}(s)=
       35.9 +0.316\  \log^2{s\over s_{0}}   
        +  42.1  \left({s\over s_{1}}\right)^{-0.468}
         -  32.2  \left({s\over s_{1}}\right)^{-0.540},
\ea
 with all coefficients in mb, $s_{1}=1$~GeV$^2$ and
 $s_{0}=34.41$~GeV$^2$.
The analysis of \cite{karol} and the data of \cite{murthy}
show that the ratio $R_{C/p}$ 
decreases very slowly in the region $5 \leq p_L \leq  600 $~GeV/$c$.
So, we take  
\ba R_{C/p}(s) = 9.5 \ (1 - 0.015 \ln{s})\ea for  the energy region
  $24 \leq p_L \leq 250 \ $GeV$/c$.

The experiment dta \cite{selex} on $pC$ scattering 
 at $p_L = 600$~GeV/$c$ 
gives us 
 $ B^+_{pC}(t \approx 0.02{\ \rm GeV}^2) =  62\ {\rm GeV}^{-2}$.
But the experiment data \cite{shiz} on hadron-nucleus scattering gives
 at small $t$
  $ B^+_{pC}(t \approx 0.01{\ \rm GeV}^2) =  70.5 \ {\rm GeV}^{-2} $
 at $p_L = 70 \ $GeV$/c$,    and 
  $ B^+_{pC}(t \approx 0.01{\ \rm GeV}^2) =  74\ {\rm GeV}^{-2} $
  at $p_L = 175 \ $GeV$/c$.
Hence, we assume that the slope slowly rises with $\ln{s}$ in a way 
similar to the $pp$ case, and normalise it 
 so that it reproduces \cite{shiz} or \cite{selex}. 
%(\ref{datasel}) or (\ref{datashi1}-\ref{datashi2}).
 These two normalisations give us a slope of similar value
  $B^{+}_{pC} \simeq 5.5 \ B_{pp}$.

We can now parameterize the spin-flip part of $p ^{12}C$ scattering as    
\ba
{\A^{h}_{sf}}(s,t)&=&   (k_2 \ + \ i \  k_1) 
  { \sqrt{|t|} \ \sigma_{tot}^{pA}(s)\over 4\pi} 
\exp\left({ B^{-}\over 2}t\right)
%\nonumber,  
\ea     
 According to the above analysis, we investigate  two extreme cases for the  
 slope of the spin-flip amplitude: \begin{itemize}
\item case I -
the spin-flip and the spin-non-flip amplitude have the same slope 
 $B^{-}_{pC} \ = \ B^{+}_{pC}$; \item 
case II - $B^{-}_{pC} \ = \ 2 B^{+}_{pC}$.\end{itemize}

One could of course allow for more freedom and try to determine
  the values of the slope $B^-$,
but the data are not yet precise enough to do this.
 The coefficients $k_1$ and $k_2$ are  chosen to obtain the best description
of $A_N$ at  $  p_L  =  24$ and $100 $~GeV/$c$. 
Of course, we can only aim at a qualitative description
as the data are only preliminary and as they are
normalised to those at $p_L = 22 $~GeV/$c$ \cite{an22}.

%Fig. 2 a,b
\begin{figure}
\epsfysize=6.cm
\epsfxsize=8.5cm
\vglue -1cm
\centerline{\epsfbox{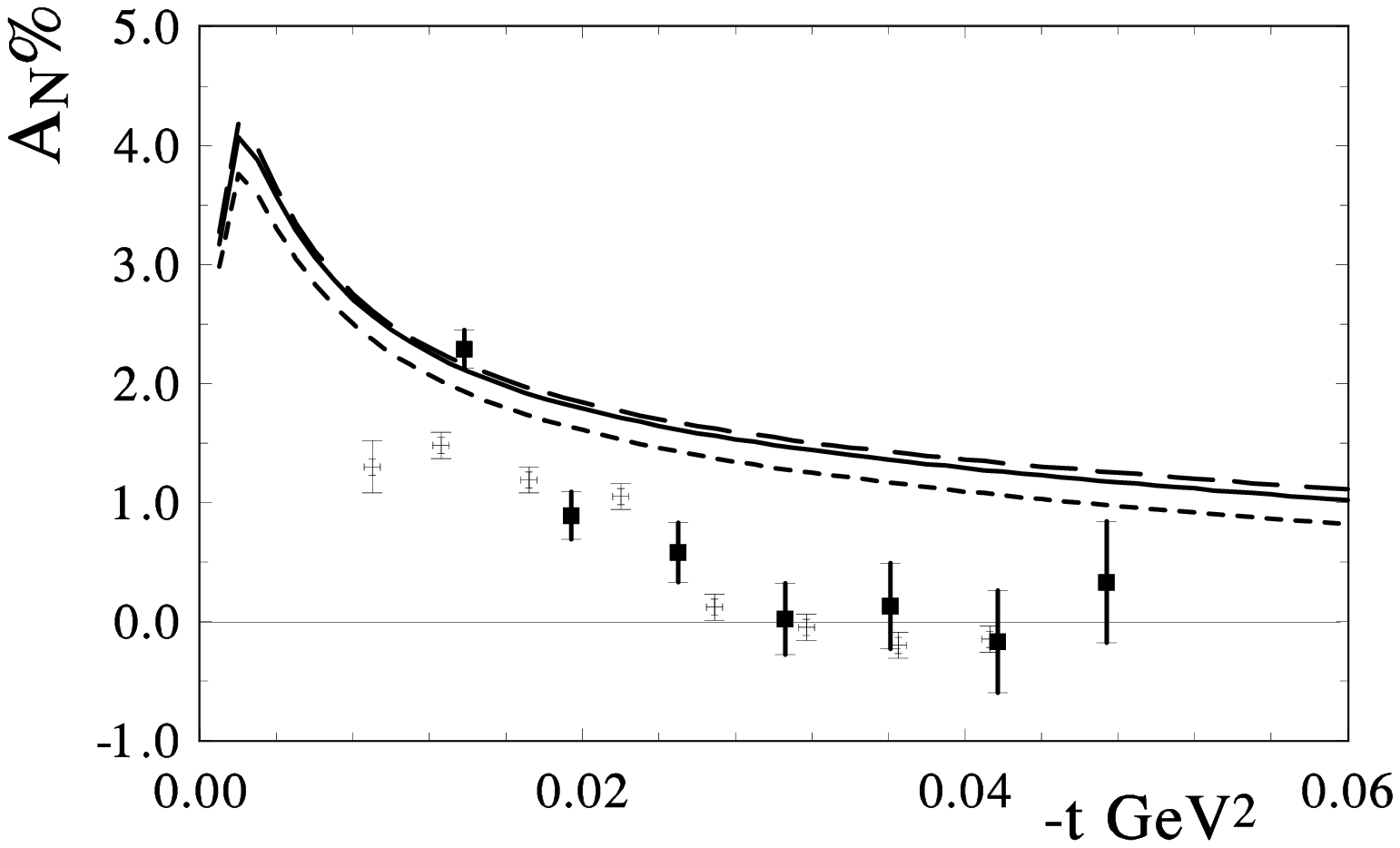}}
\vglue -1cm
\centerline{\epsfbox{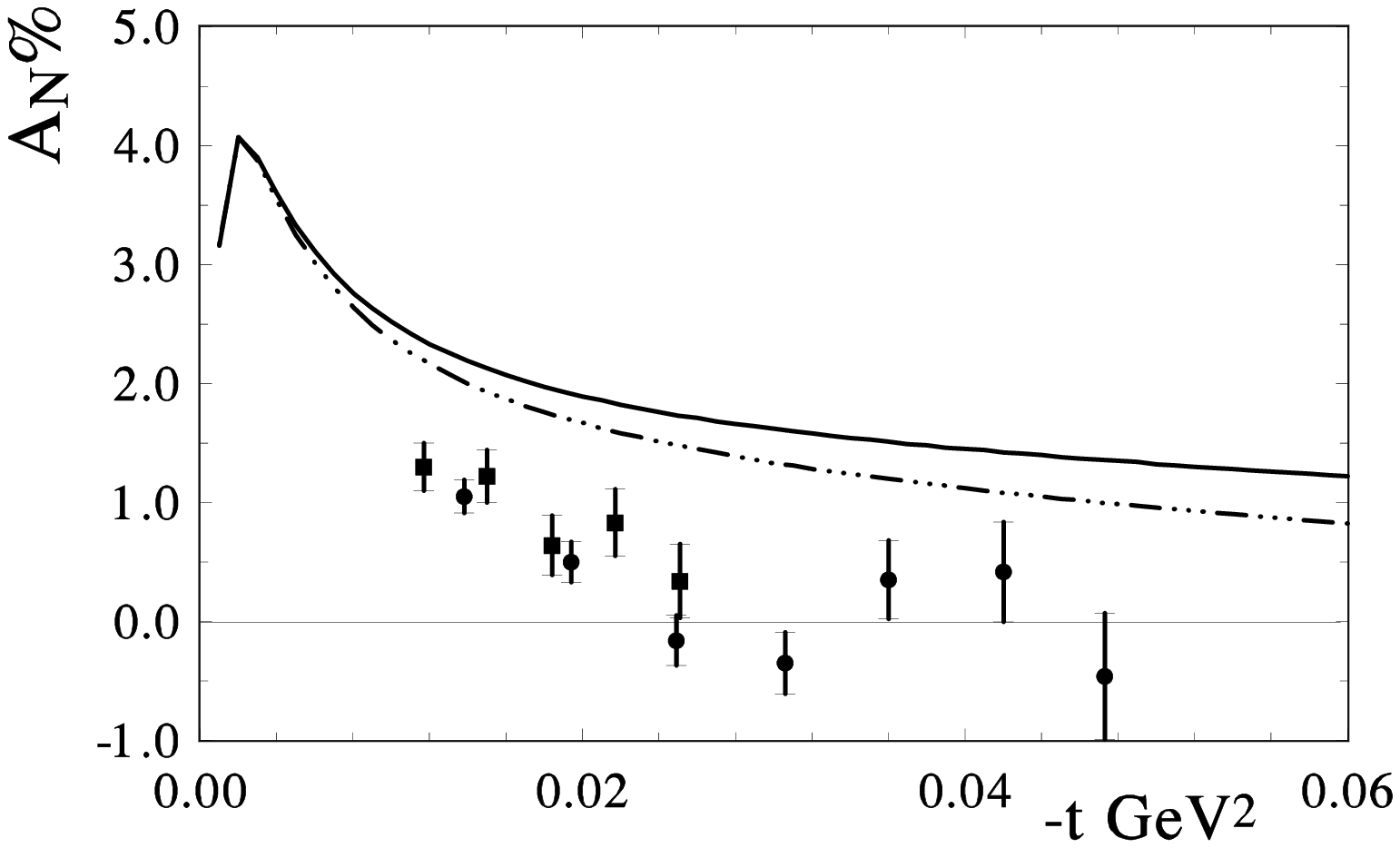}}
\caption{ \hfill\break
a)  The energy  dependence of $A_N$ (without hadron spin-flip)
at $p_{L} = 24, \ 100, \ 250  $~GeV/$c$ (dashed, solid, and long-dashed lines
  correspondingly) 
 compared with the data  at $p_{L} = 22 \ $GeV/$c$ (crosses) and 
   $p_{L} = 24 \  $~GeV/$c$ (boxes)
\cite{an22,an100} (only statistical errors are shown). 
\hfill\break b) The dependence of $ A_N$ (without hadron spin-flip)
            over $B^+$ (solid line -  $B^{+}$ 
is normalised to the data of \cite{shiz}; dash-dotted line - $B^{+}$ is
            normalised to the data of \cite{selex})
compared with the data  at $p_{L} = 100 \ $GeV/$c$ (boxes and circles) 
\cite{an100} (only statistical errors are shown) }

\end{figure}

>From the full scattering amplitude, the analysing power is given by  
\ba
  A_N\frac{d\sigma}{dt} =
         - 4 \pi [Im(\A_{nf})Re(\A_{sf})-Re(\A_{nf})Im(\A_{sf})],
%\nonumber 
\ea 
each term having electromagnetic and hadronic contributions.

  Now, let us first calculate  $A_N$ without the contribution of 
 the hadron-spin-flip amplitude. In this case, the size and shape of  $A_N$
 is determined by the interference between the hadron spin-non-flip
 amplitude and the magnetic part of the electromagnetic amplitude.
  In Fig.~2 (a), the results of such calculations are shown for    
small momentum transfer, at $p_{L} = 24, \ 100, \ 250  $~GeV/$c$, 
and are compared with the preliminary data  at $p_L = 22, \ 24 \ $ GeV/$c$ 
  \cite{an22,an100}.  The energy dependence is weak and 
  is determined in mostly by the energy dependence of $\rho$, which we
  have taken proportional to  $\rho_{pp}$.
   In Fig.~2 (b), the calculations are made at $p_{L} =  \ 100 \  $~GeV/$c$
  for the different normalisations of the slope.
% (\ref{datasel}) and (\ref{datashi1}-\ref{datashi2}).
  These lead respectively, at $p_{L} =  \ 100 \  $~GeV/$c$,
 to the values
  $B_{pC}^{+} = 58.3 \ $GeV$^{-2}$ and  $B_{pC}^{+} = 72.1 \ $GeV$^{-2}$.
  It is clear that this difference only slightly changes the size of $A_N$ at
  $|t| \geq 0.02 \ $GeV$^2$.

%Fig. 3 a,b
\begin{figure}
\epsfysize=6.cm
\epsfxsize=8.5cm
\vglue -1cm
\centerline{\epsfbox{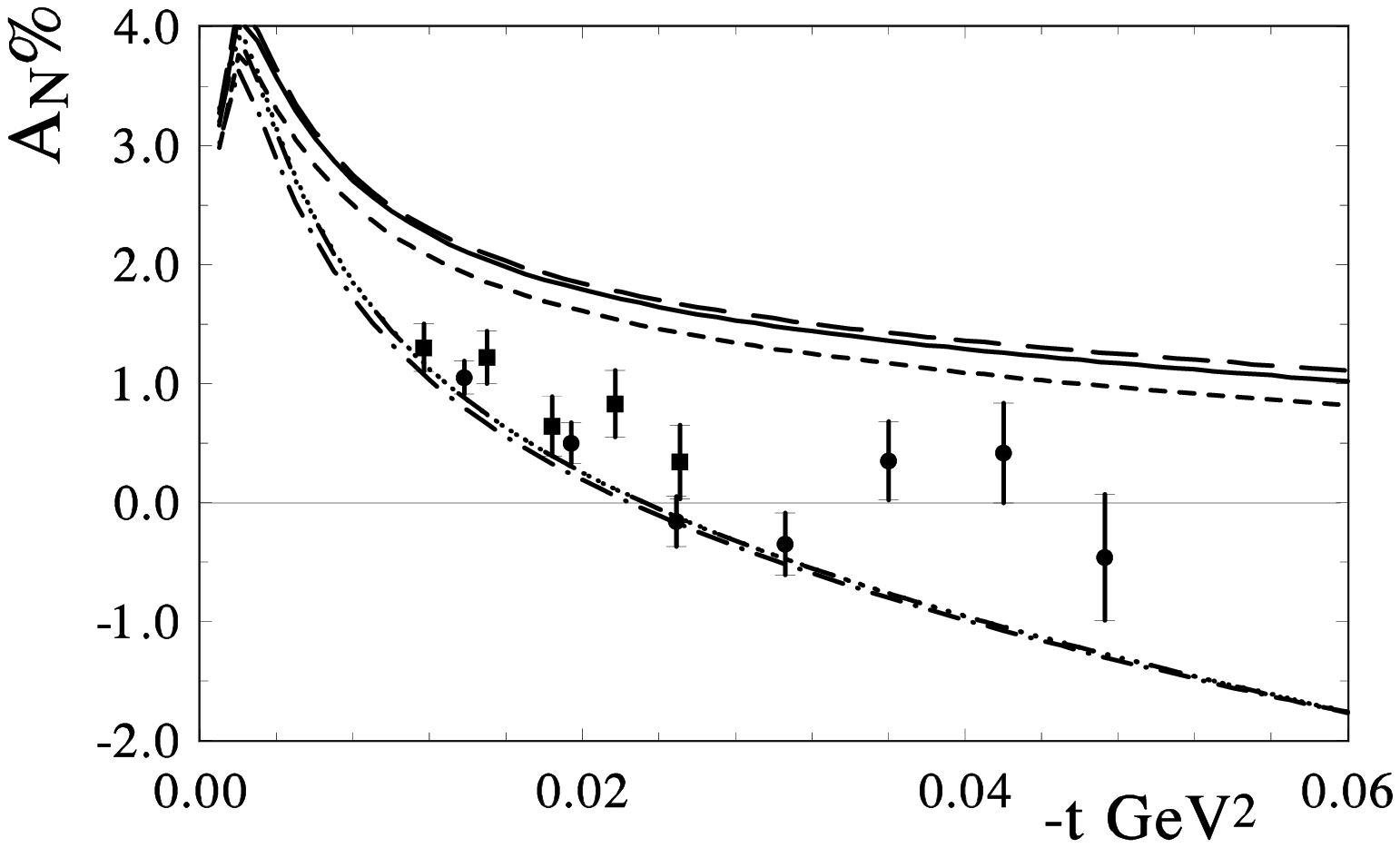}}
\vglue -1cm
\centerline{\epsfbox{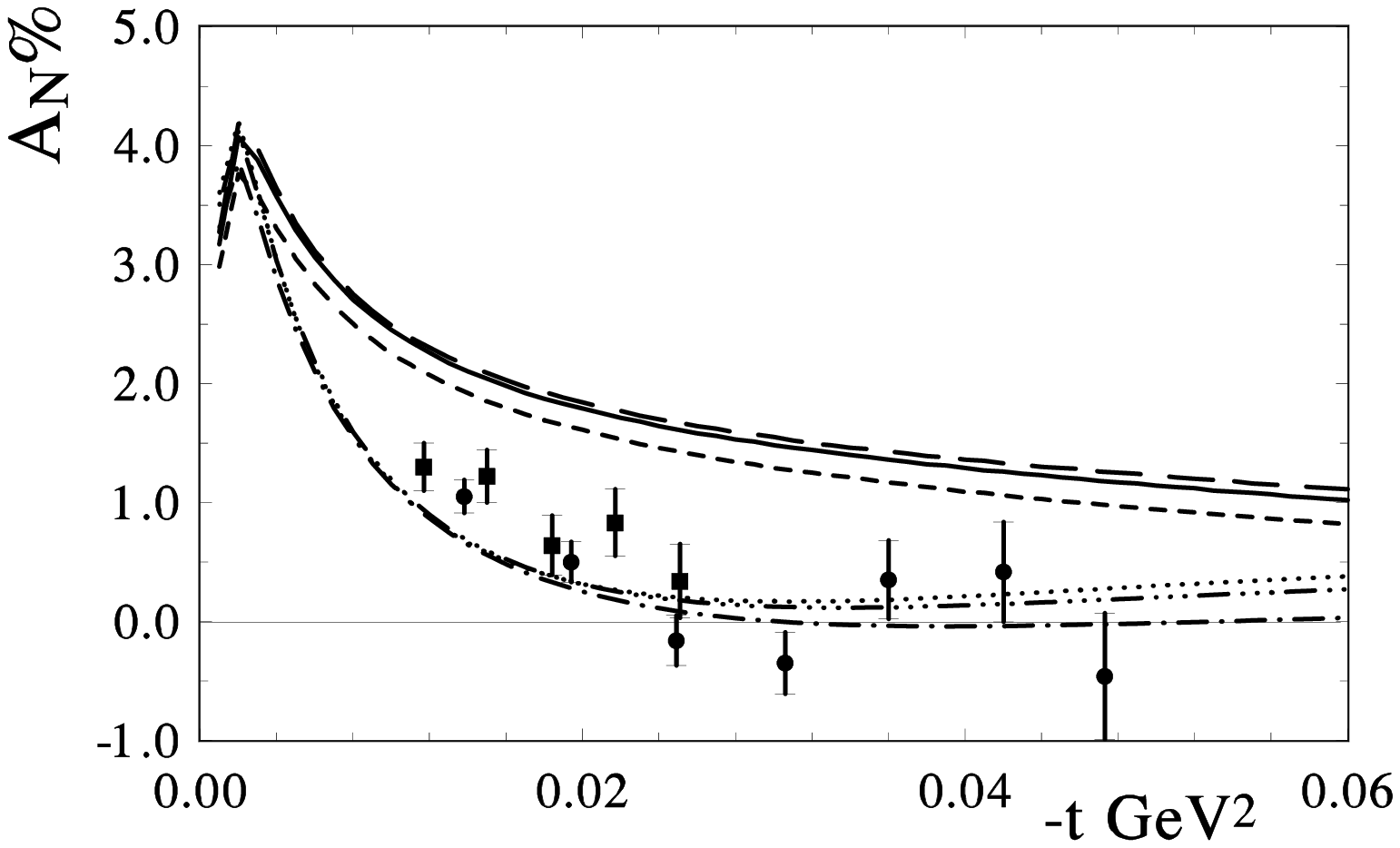}}
\caption{ \hfill\break
a)   $A_N$ with hadron-spin-flip amplitude in 
          case I ($B^{-} = B^{+}$) for $p_L = 24, \ 100, \   250 \  $GeV/$c$
          (dash-dot, dash-dots, and dots  correspondingly), 
 compared with the curves of Fig.~2 (a).\hfill\break
   b) $A_N$ with hadron-spin-flip amplitude in 
          case II ($B^{-} =2 B^{+}$) for 
      $ p_L = 24,  \ 100, \ 250 \ $GeV/$c$.
       (dash-dot, dash-dots, and dots  correspondingly), 
 compared with the curves of Fig.~2 (a).}
\vspace{5mm}
\end{figure}

  In Fig.~3, we show  the values of $A_N$ calculated for cases I and II
  for the slope of the hadron spin-flip amplitude.
  We can see that in both cases we obtain a small energy dependence.
  In case I, $A_N$ decreases less with $|t|$ immediately 
  after the maximum. But at large $|t| \geq 0.01$~GeV$^2$ the behaviour of
  $A_N$ is very different: we can  obtain a different sign for $A_N$ at
  $|t| \approx 0.06$~GeV$^2$. 
   In  case I, when $B^{-}_{pC} = B^{+}_{pC}$,   $A_N$ changes sign 
  in the region of $|t| \approx 0.02$~GeV$^2$ and then grows in magnitude.
In case II,   when $B^{-}_{pC} = 2 B^{+}_{pC}$,  $A_N$ approaches zero
   and then grows  positive again. 

  It is interesting to note that in \cite{sc}, 
  where we investigated the analysing power for $p ^{12}C$-reaction 
  in case I, but with  a more complicated form factor,
  we again  obtained the possibility that
   the slope  of the hadron spin-flip  exceeds the value
  $60\ $GeV$^{-2}$, and we showed that both slopes at very small momentum
  transfer were  equal to about  $90\ $GeV$^{-2}$.  
  Of course, such a large slope 
  for the spin-non-flip amplitude 
  cannot be obtained in the standard Glauber approach and would have to come
from another mechanism.  
      
   The preliminary experimental data show that $A_N$ decreases
  very fast after its maximum and is almost zero in a small
  region of momentum transfer. This behaviour can be explained
  only if one assumes a negative contribution of the interference 
  between different parts of the hadron amplitude, which    
  changes slowly with energy. 
  The preliminary data at  $p_L = 100 $~GeV/$c$ decrease faster
  than those at $p_L = 24 $~GeV/$c$, and  the zero 
  of $A_N$ may move to lower values of $|t|$.
  This change of sign is independent from the normalisation 
  of the data. It would be very interesting to
  obtain new data with higher accuracy and at higher energies in order to
  distinguish between the two scenarios for the slopes of the spin-flip
  amplitude.

   Of course, the weak energy dependence 
  comes from our choice of the energy dependence of the spin-flip amplitude.
  If we introduce {\it e.g.} an additional 
  factor $\sqrt{s_1/s}$ with $s_1 =6.83 \ $GeV 
  (that $s_1$ corresponds to $p_L=24 \ $GeV$/c$), 
  we obtain a strong energy dependence for $A_N$, 
  which is shown in Figs.~4 (a,b) in both cases,
  but such an energy dependence contradicts the existing experimental data.  
  If we made the coefficient $k_2$ proportional to the $\rho(s)$
  of the spin-non-flip amplitude, 
   we again would obtain a strong energy dependence
  which is shown in Figs.~5 (a,b). 
  Hence, we can conclude that already for the present energy 
  of proton-nucleus scattering,   the part of 
  the hadron-nucleus  spin-flip amplitude connected with secondary Reggeons
  is small and there exists a non-negligible  part 
   of the hadron-nucleus spin-flip amplitude 
  in which the imaginary and real parts have  a weak energy dependence.    

%Fig. 4 a,b
\begin{figure}
\epsfysize=6.cm
\epsfxsize=8.5cm
\vglue -1cm
\centerline{\epsfbox{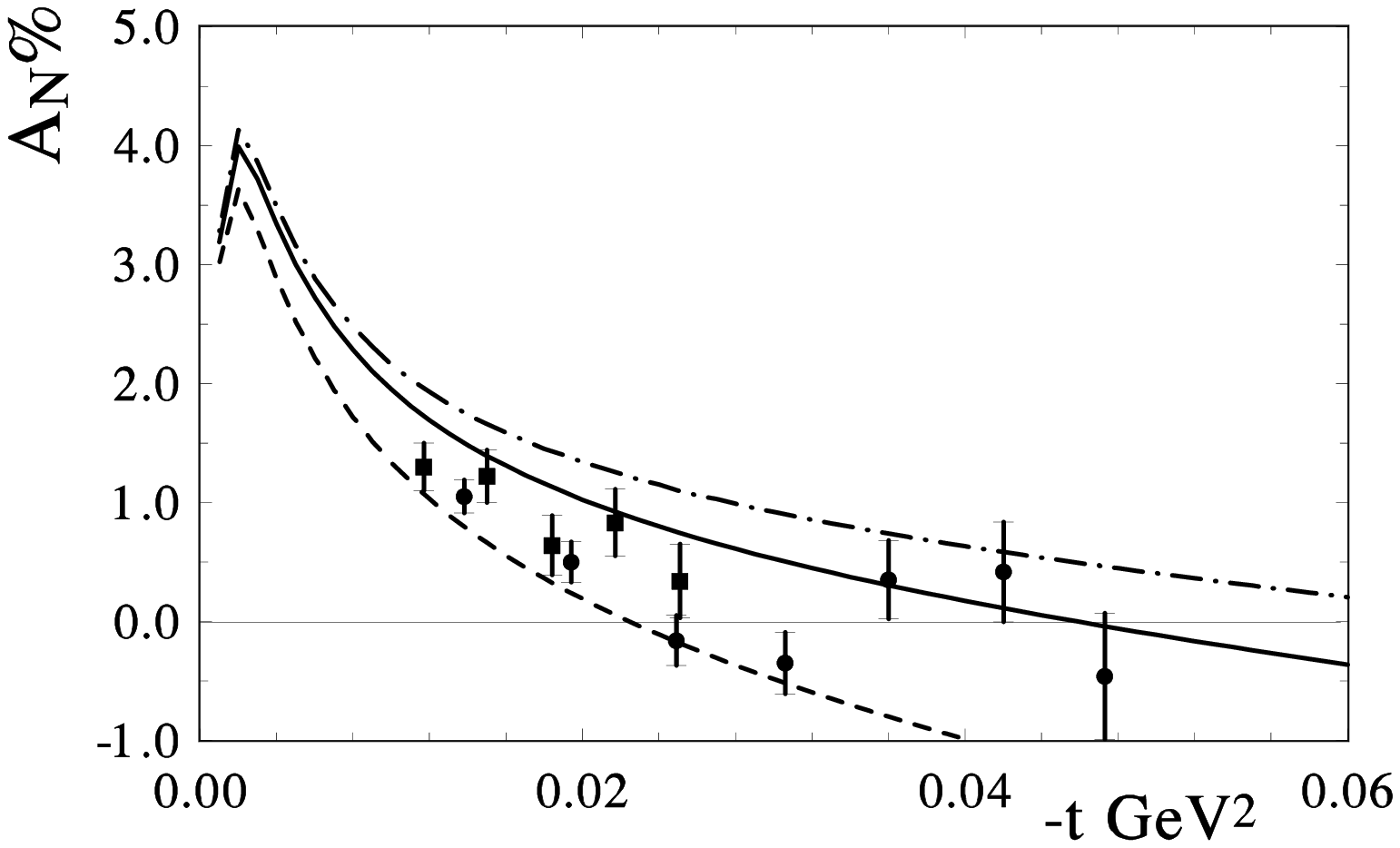}}
\vglue -1cm
\centerline{\epsfbox{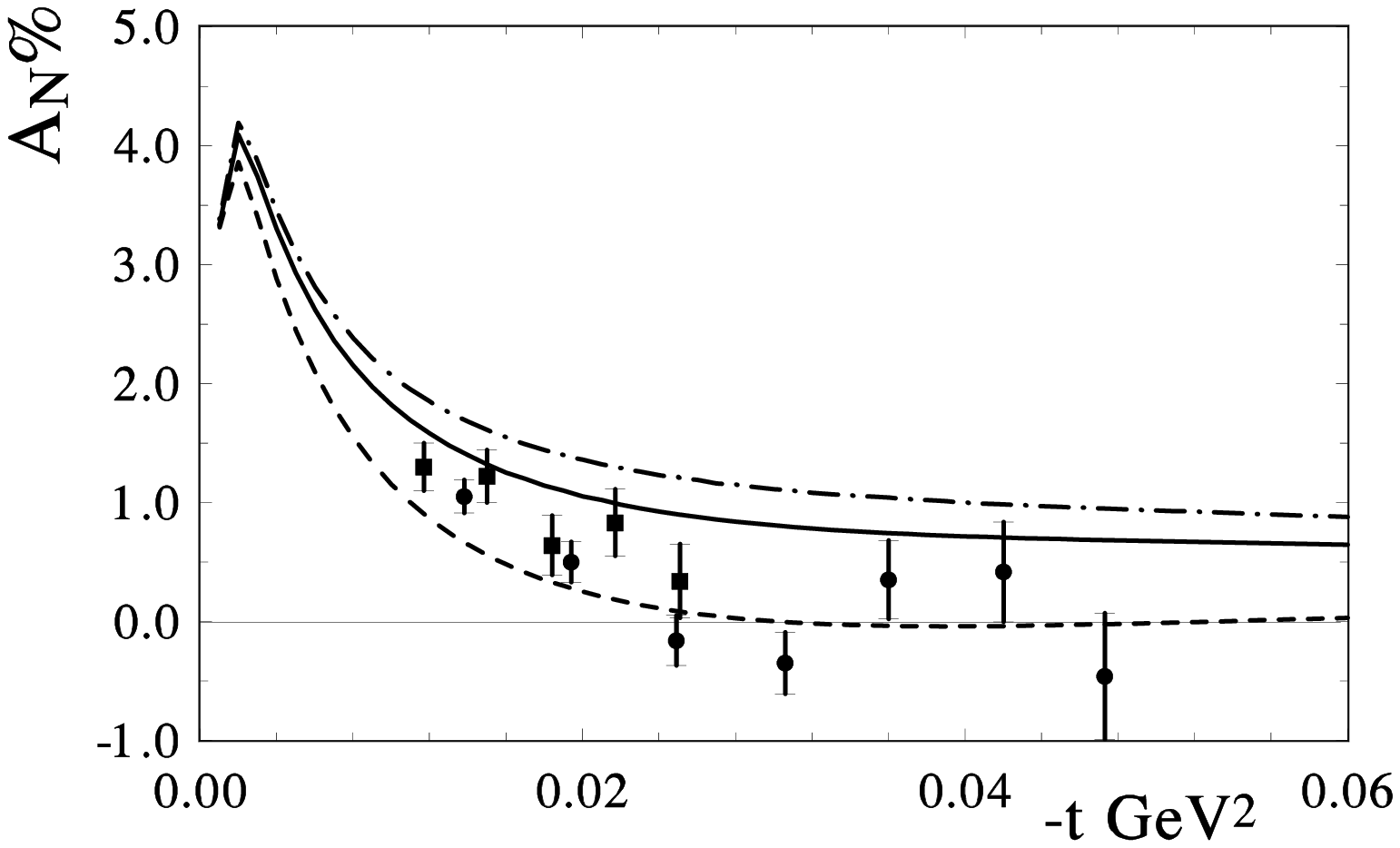}}
\caption{   $A_N$ with a hadron spin-flip amplitude which has
           an additional  energy factor $\sqrt{s_1}/\sqrt{s}$ 
    \hfill\break
a) in  case I ($B^{-} = B^{+}$) for $p_L = 24, 100,  250 \  $GeV/$c$.
          (dash, solid, and dash-dot  correspondingly).
  \hfill\break
   b)the same   in  case II ($B^{-} =2 B^{+}$).
%  for       $ p_L = 24,  \ 100, \ 250 \ $GeV/$c$.
%       (dash-dot, dash-dots, and dots correspondingly).
}
\vspace{5mm}
\end{figure}

%Fig. 5 a,b
\begin{figure}
\epsfysize=6.cm
\epsfxsize=8.5cm
\vglue -1cm
\centerline{\epsfbox{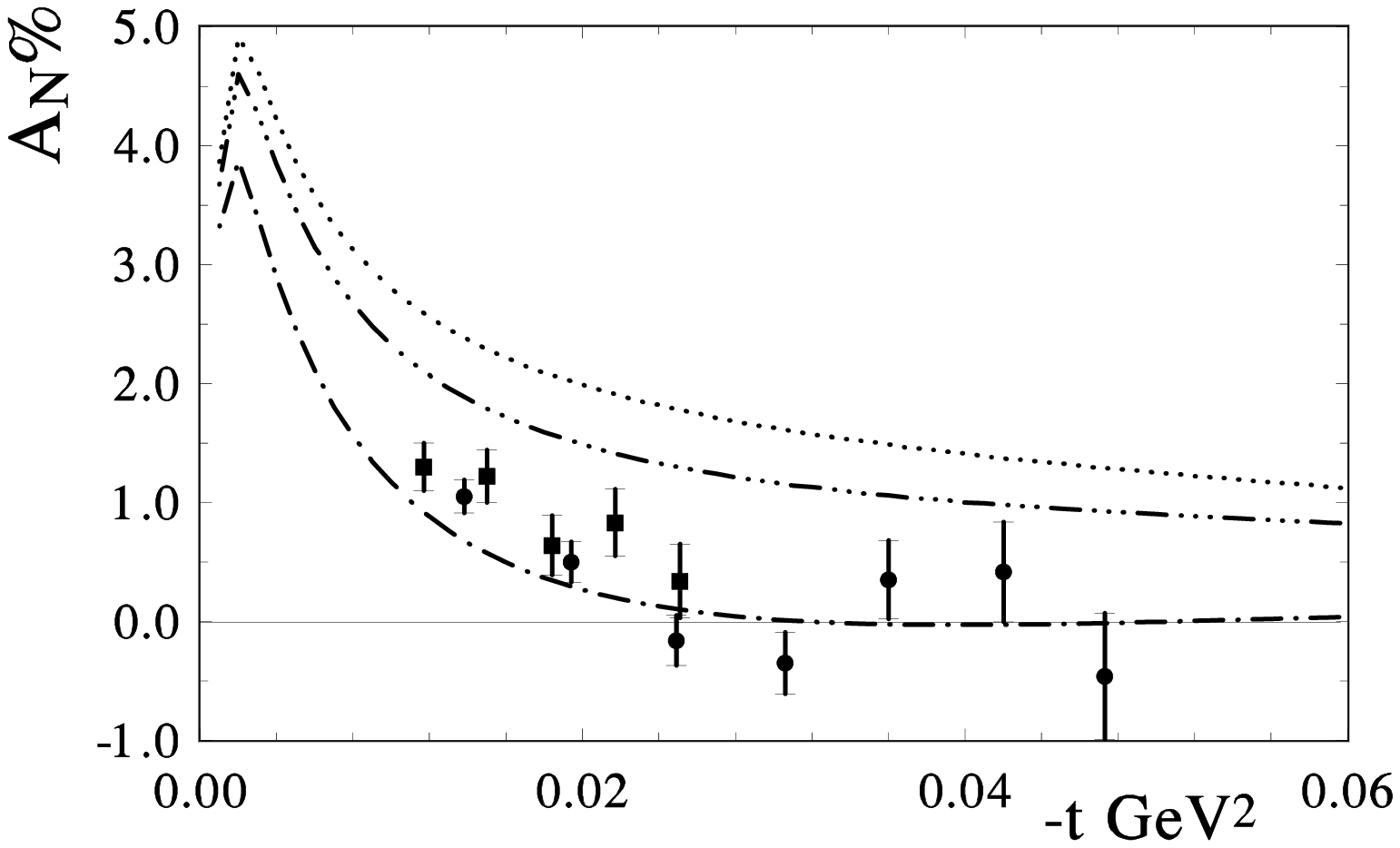}}
\vglue -1cm
\centerline{\epsfbox{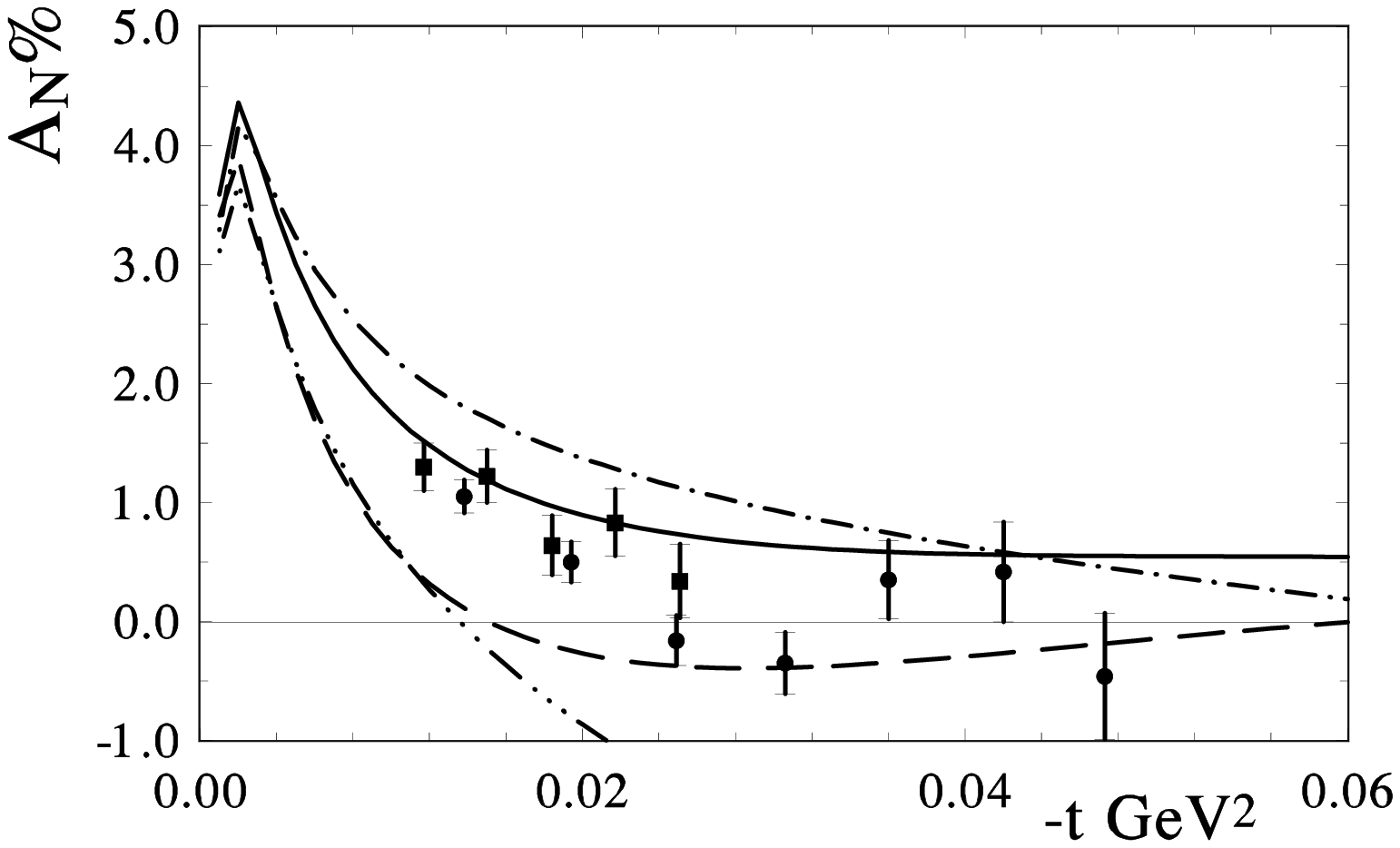}}
\caption{ \hfill\break
a)  $A_N$    in  case II ($B^{-} \ = \ 2B^{+}$)      
 and with the real part of the spin-flip
   amplitude proportional to $\rho$ at 
 $ p_L = 24,  \ 100, \ $ and $250 \ $GeV/$c$. 
       (dash-dot, dash-dots, and dots  correspondingly).\hfill\break
   b)  The values of $A_N$ corresponding to a change of $\pm 0.04$ in $k_2$
      in  cases (I) and (II)
   ( dash-dots and dash-dot are for the case I;
       the  dashed and solid lines are for case II)
 for   $ p_L = \ 100 \  $GeV/$c$. }
\vglue 0.2cm
\end{figure}

\section{Conclusion}

  By  accurate measurements of the analysing power in
   the  Coulomb-hadron
   interference region, we can find the structure of the hadron spin-flip
   amplitude, and this obtain further information about the
  behaviour of the hadron interaction potential at large distances.
   Contributions beyond the usual eikonal formalism are expected
in the peripheral
  dynamic model \ci{zpc}, which takes into account 
hadron-hadron
 interactions at large  distances. The resulting 
 ``residual'' hadron spin-flip amplitude
 has a different slope from that of the
  spin-non-flip amplitude at small momentum transfer.
 The model also gives  large spin effects in the diffraction dip
  region \cite{yaf-str}.
  We should note that all our consideration are based
  on the usual assumptions that
  the imaginary part of the high-energy scattering amplitude
  has an exponential behaviour.
  The other possibility, that  the slope changes slowly as
    $ t \rightarrow 0 $, requires a more refined discussion
   that will be the subject of a subsequent paper.
 
\bigskip

\noindent {\it Acknowledgements} O.V.S. is a Visiting Fellow of the 
Fonds National pour la 
Recherche Scientifique, Belgium. We thank V.~Kanavets and D.~Svirida
for their comments and discussions.
%\end{acknowledgments}  

\end{document}